\documentclass[aps,onecolumn,nofootinbib,preprintnumbers,superscriptaddress]{revtex4}
\pdfoutput=1

\usepackage{tabu}
\usepackage[usenames,dvipsnames,table]{xcolor}
\usepackage{graphicx,amsmath,amssymb,amsthm,multirow,array,bm,bbm,esint}
\usepackage[mathscr]{eucal}
\usepackage[bbgreekl]{mathbbol}
\usepackage{epsf,amsfonts}
\usepackage{slashed}
\usepackage{hyperref}

\usepackage{ulem}

\hypersetup{
    pdfstartview={FitH},    
    pdftitle={Entropy Inflow},    
    pdfauthor={Felix Haehl, R.~Loganayagam, Mukund Rangamani},     
    colorlinks=true,       
    linkcolor=blue,          
    citecolor=red,        
    filecolor=magenta,      
    urlcolor=blue           
}
\definecolor{rust}{rgb}{0.8,0.2,0.2}

\newcommand{\Dut}{\SF{\mathcal{D}}}
\newcommand{\z}{\mathbf{z}}
\newcommand{\zsf}{\SF{\z}}

\newcommand{\Dwv}{\SF{\mathfrak{D}}}
\newcommand{\Cwv}{\SF{\mathscr{C}}}

\newcommand{\gpsi}{\SF{\mathfrak{g}}^{(\psi)}}
\newcommand{\gpsib}{\SF{\mathfrak{g}}^{(\psib)}}

\definecolor{rust}{rgb}{0.8,0.2,0.2}

\newcommand{\prn}[1]{\left ( #1 \right )}

\newcommand{\vev}[1]{\langle #1 \rangle}

\newcommand{\Q}{\mathcal{Q}}
\newcommand{\Qb}{\overline{\mathcal{Q}}}

\newcommand{\thb}{{\bar{\theta}} }
\newcommand{\thetab}{\bar{\theta} }
\newcommand{\SF}[1]{\mathring{#1}}

\newcommand{\LamS}{\SF{\Lambda}}
\newcommand{\As}{\SF{\mathscr{A}}}
\newcommand{\Ascr}{\mathscr{A}}

\newcommand{\Fs}{\SF{\mathscr{F}}}

\newcommand{\psib}{\overline{\psi}}

\newcommand{\tx}{\tilde{X}}
\newcommand{\xpsi}{X_\psi}
\newcommand{\xpsib}{X_{\psib}}

\def\source#1{{\textcolor{black}{ #1}}}

\def\sBdel{\source{{\sf B}_\Delta}}
\def\sAt#1{\source{\mathcal{F}_{#1}}}

\newcommand{\Kref}{{\bm \beta}}

\newcommand{\gref}{{\sf g}}

\newcommand{\Lagref}{\mathscr L}

\newcommand{\etaref}{\bm \eta}

\newcommand{\Nref}{{\sf \mathbf N}}
\makeatletter
  \newcommand\Ttiny{\@setfontsize\Ttiny{1pt}{2}}
\makeatother

\newcommand{\TEMref}{{\sf\mathbf T}}

\newcommand{\Cref}{{\sf C}}

\newcommand{\lieD}{\pounds}

\newcommand{\UT}{U(1)_{\scriptstyle{\sf T}}}

\begin{document}

\title
{An Inflow Mechanism for Hydrodynamic Entropy
}

\author{Felix M. Haehl}
\email{f.m.haehl@gmail.com}
\affiliation{Department of Physics and Astronomy, University of British Columbia, 
6224 Agricultural Road, Vancouver, B.C. V6T 1Z1, Canada.}

\author{ R.\ Loganayagam}
\email{nayagam@gmail.com}
\affiliation{International Centre for Theoretical Sciences (ICTS-TIFR),
Shivakote, Hesaraghatta Hobli, Bengaluru 560089, India.}

\author{Mukund Rangamani}
\email{mukund@physics.ucdavis.edu}
\affiliation{Center for Quantum Mathematics and Physics (QMAP)
\& Department of Physics, University of California, Davis, CA 95616 USA.}

\begin{abstract}
We argue that entropy production in hydrodynamics can be understood via a superspace inflow mechanism. Our arguments are based on a recently developed formalism for constructing effective actions for Schwinger-Keldysh observables in quantum field theories. The formalism explicitly incorporates  microscopic unitarity and the  Kubo-Martin-Schwinger thermal periodicity conditions, by recasting them in terms of topological BRST symmetries of the effective action.

\end{abstract}


\maketitle

\section{Introduction}
\label{sec:intro}

Non-linear hydrodynamics, which is the long-wavelength, low-energy theory of near-equilibrium systems, is traditionally characterized by the following set of axioms \cite{landau}:\footnote{We phrase our discussion for a neutral relativistic fluid. Other conserved charges can be included in a straightforward manner.}
\begin{itemize}
\item[H1.] The low energy variables are the fluid velocity $u^\mu$ ($u^\mu\,u_\mu =-1$), and intensive thermodynamic parameters such as temperature $T$.  We combine these into the thermal vector $\Kref^\mu = u^\mu/T$. 
\item[H2.] The currents are functionals of fluid variables and background sources: $T^{\mu\nu}[u^\mu, T, g_{\mu\nu}]$. They admit a gradient expansion with each independent tensor structure defining a potential transport coefficient.
\item[H3.] Dynamics is contained in conservation laws: $\nabla_\mu T^{\mu\nu} =0$ for energy-momentum. 
\item[H4.] Admissible constitutive relations require the existence of an \emph{entropy current} $J^\mu_S$, which satisfies a local form of the second law of thermodynamics, $\nabla_\mu J^\mu_S = \Delta \geq 0$, for every physical fluid configuration. 
\end{itemize}

Whilst it is easy to intuit that the low-energy theory is characterized by conservation laws, as  only the corresponding modes survive to dominate the late-time, long-distance behaviour, the entropy current is more mysterious. For one, it emerges in the infra-red without being manifest in the ultra-violet, and appears in gravitational systems to be associated with black hole horizons.  This stark dichotomy can be seen in the context of the fluid/gravity correspondence \cite{Hubeny:2011hd} where the conserved currents live on the boundary of the AdS spacetime, while the entropy current is obtained from the interior by pulling back the area form of the horizon 
\cite{Bhattacharyya:2008xc}. 

Given this status quo, one can ask, whether there is a more intuitive way to understand the origin of the hydrodynamic entropy current, $J^\mu_S$, and its associated entropy production $\Delta$? We argue here for the following statements: 
\begin{itemize}
\item For systems in local equilibrium the macroscopic free energy current, the Legendre transform of entropy current, $N^\mu = J^\mu_S + \Kref_\nu T^{\mu \nu}$, is a Noether current for thermal diffeomorphisms along $\Kref^\mu$. 
\item The hydrodynamic effective theory is one where these thermal diffeomorphisms are gauged as a topological/BRST gauge symmetry, leading to the idea of \emph{thermal equivariance}. This symmetry can be made manifest by formulating the theory in superspace -- an extension of ordinary spacetime by two anti-commuting directions.
\item While the net entropy being conjugate to the gauged thermal diffeomorphisms is conserved, entropy in physical space is produced by virtue of it being sourced in the superspace directions, i.e., there is an inflow of entropy from superspace.  This inflow is non-zero in when appropriate superspace components of the field strength are turned on.
\end{itemize}
The mechanism for entropy production is thus analogous to the picture for 't Hooft anomalies in field theories. Recall that the inflow mechanism \cite{Callan:1984sa} implies that a theory with a 't Hooft anomaly can be coupled to a higher-dimensional topological field theory so as to leave the combined system anomaly free. In the presence of background field strengths, the physical theory acquires its anomaly through an inflow from the topological sector. In the present context, the symmetry under question is thermal diffeomorphisms, which involve taking fields around a fiducial Euclidean circle, and the inflow is from the topological sector of this theory, which has been recast into a superspace description. An appropriate superspace field strength, drives this `entropy anomaly' in the physical space.

\section{Hydrodynamic effective field theories}
\label{sec:heff}

Any interacting QFT will attain thermal equilibrium in a stationary spacetime with a timelike Killing field $K^\mu$ \cite{Banerjee:2012iz,Jensen:2012jh}. Equivalently, equilibrium questions in thermal QFT, can be answered by a statistical field theory formulated on a Riemannian manifold $\mathcal{M}_\Kref$ with the Euclidean thermal circle fibered over a spatial base. The generating functional for current correlators is the equilibrium partition function on this non-trivial background. Such configurations are hydrostatic: the fluid velocity points along the Killing field and the local temperature is given by its norm,  $\Kref^\mu = K^\mu$. The free energy current for the fluid is precisely given by the Noether current for diffeomorphisms along the thermal circle \cite{Haehl:2014zda}, a statement that may be intuitively understood by recalling the fact that stationary black hole entropy is also obtained as a Noether charge \cite{Iyer:1994ys}.

Beyond equilibrium one has to confront challenges of defining entropy, free energy etc. However, Bhattacharyya's theorem \cite{Bhattacharyya:2014bha} states that a local free energy current exists in near-equilibrium hydrodynamic settings satisfying a local form of the second law, as long as the  hydrostatic limit ($\Kref^\mu \rightarrow K^\mu$) is consistent with the equilibrium partition function. Remarkably, the second law of thermodynamics only requires in addition that leading order dissipative terms are sign-definite (e.g., viscosities 
\& conductivities are non-negative). More generally, as argued by us in \cite{Haehl:2014zda,Haehl:2015pja}, hydrodynamic transport coefficients can be classified into 8 distinct classes by examining the off-shell entropy production statement embodied in the \emph{adiabaticity equation}:
\begin{equation}
\nabla_\mu N^\mu - \frac{1}{2}\, T^{\mu\nu} \lieD_\Kref g_{\mu\nu} = \Delta \geq 0 \,.
 \label{eq:ordadiab}
\end{equation}	
Dissipative solutions to this equation comprise a single class, with the remaining 7 being adiabatic or non-dissipative (i.e., $\Delta=0$). The dissipative constitutive relations are characterized by the most general symmetric 4-tensor built from the fluid data, 
\begin{equation}
\begin{split}
T^{\mu\nu}_{diss} &= \frac{1}{2}\, {\bm \eta}^{(\mu\nu)(\rho\sigma)} \, \lieD_\Kref g_{\rho \sigma} \,, \quad\; {\bm \eta}^{(\mu\nu)(\rho\sigma)}={\bm \eta}^{(\rho\sigma)(\mu\nu)} \,,
\end{split}
 \label{eq:Tdiss}
\end{equation}	
which provides a non-negative definite inner product on the space of symmetric two tensors:
\begin{equation}
\Delta = \frac{1}{4}\, {\bm \eta}^{(\mu\nu)(\rho\sigma)}\,\lieD_\Kref g_{\mu\nu}\, \lieD_\Kref g_{\rho \sigma} \geq 0\,.
\end{equation}
The curiosity of this classification is that the remaining 7 classes of transport  do not lead to entropy production -- fluid transport with $\Delta = 0$ is surprisingly rich. Given that the adiabaticity equation is off-shell, i.e., fluid conservation equations are not imposed, one should ask for a general principle that explains their presence beyond hydrostatics.

Based on empirical observations justified post-facto by the structure of Ward identities in   Schwinger-Keldysh (SK) functional integrals \cite{Schwinger:1960qe,Keldysh:1964ud} which capture response functions, and the Kubo-Martin-Schwinger (KMS) relations \cite{Kubo:1957mj,Martin:1959jp} which enforce the fluctuation-dissipation theorem, several groups have argued that the hydrodynamic effective field theories should be constrained by  topological BRST symmetries \cite{Kovtun:2014hpa,Haehl:2015foa,Crossley:2015evo,Haehl:2015uoc,Haehl:2016pec,Haehl:2016uah,Glorioso:2016gsa,Glorioso:2017fpd,Gao:2017bqf,Jensen:2017kzi,Haehl:2017zac}.\footnote{ Note that such a structure may be naturally anticipated given the BRST symmetry inherent in effective actions for Langevin systems and stochastic field theories \cite{Martin:1973zz,Parisi:1982ud}.}

Hydrodynamic transport is captured by response functions, which are causal Green's functions, obtained by starting with a system in global equilibrium, disturbing it by turning on a sequence of time-dependent background sources, and finally making a measurement of the effect. In the SK formalism phrased in the Keldysh average-difference basis, response functions are correlators of a sequence of difference operators (the disturbances) followed by an average operator (the measurement) in the future. This is the first non-vanishing correlation function, since pure difference operator correlators are constrained to vanish by unitarity \cite{Chou:1984es}. Furthermore, the response functions are related by the KMS conditions to correlators with multiple average operators; the latter capture fluctuations, a subset of which are hydrodynamic.\footnote{ Understanding the KMS constraints on higher-point functions requires time contours beyond the SK formalism \cite{Haehl:2017eob}.}

The BRST symmetries constrain the low energy dynamics by picking out the correct influence functionals \cite{Feynman:1963fq} which arise from integrating out the high-energy modes. The macroscopic theory has two sets of degrees of freedom: the average or classical fields, and the difference or quantum/stochastic fluctuation fields. The consistent couplings are dictated by the ghost degrees of freedom which form part of the BRST multiplet. All of this information can be succinctly encoded by working in a superspace that is locally ${\mathbb R}^{d-1,1|2}$ with coordinates $z^I = \{\sigma^a,\theta, \thb\}$.

While the superspace description encodes SK unitarity constraints, imposing the KMS conditions requires additional structure which may be phrased in the framework of equivariant cohomology
\cite{Haehl:2016pec,Haehl:2016uah}. The idea is to imagine out-of-equilibrium systems in real time, as possessing a local thermal circle fibred over the Lorentzian base space. This fibration is captured by a background timelike vector superfield  $\SF{\Kref}^I(z)$, which is the superspace lift of the thermal vector.\footnote{ We will use some of the superdiffeomorphism invariance to simplify the thermal super-vector to $\SF{\Kref}^\theta=\SF{\Kref}^{\thetab} = 0=\partial_\theta \SF{\Kref}^a=  \partial_{\thetab} \SF{\Kref}^a  $. Note that we consistently use an over-circle to distinguish superfields from ordinary fields.} Working in the high temperature limit, one may then argue that the BRST symmetries are deformed by a thermal $\UT$ gauge field, which implements invariance under thermal diffeomorphisms. On scalar superfields, the $\UT$ symmetry transformation acts as  $\LamS' \mapsto \LamS' + (\LamS,\LamS')_\Kref $  with the bracket relation
\begin{align}
(\LamS,\LamS')_\Kref &=\LamS\, \lieD_\Kref \LamS'-\LamS'\, \lieD_\Kref \LamS \,.
\label{eq:adbetabrk}
\end{align}

All of this information can be captured by a gauge theory with gauge potentials,  covariant derivatives and field strengths given in superspace by \cite{Haehl:2016uah}:
\begin{equation}
\begin{split}
&\As_I(z)\, dz^I 
= 
	\As_a(z)\, d\sigma^a +\SF{\mathscr{A}}_{\theta}(z)\, d\theta + \SF{\mathscr{A}}_{\thetab}(z) \, d\bar{\theta} \,,\\
 &\Dut_I 
 = 	
 	\partial_I + (\As_I,\ \cdot\, )_\Kref\,, \\
 &\Fs_{IJ}  
 \equiv
 	(1-\frac{1}{2}\, \delta_{IJ}) \left(\partial_I\, \As_J - (-)^{IJ} \,\partial_J\, \As_I  + (\As_I,\As_J)_\Kref \right) .
\end{split}
\label{eq:Aform}
\end{equation}
The upshot of this encoding is that the hydrodynamic effective field theory in the high temperature limit is constrained by a BRST superalgebra \cite{Haehl:2015uoc} with Grassmann-odd supercharges $\Q, \Qb$ satisfying:
\begin{equation}
\Q^2 = -\Fs_{\thb\thb} | \lieD_\Kref \,,  \qquad 
\Qb^2 = -\Fs_{\theta\theta} | \lieD_\Kref  \,, \qquad  
\{\Q, \Qb\} = -\Fs_{\theta\thb} | \lieD_\Kref
\label{eq:skkmsalg}
\end{equation}	
where $|$ denotes restriction to $\theta=\thb =0$.
A closely related algebra has appeared in the statistical mechanics literature in the context of stochastic Langevin dynamics \cite{Mallick:2010su}. This is also the high temperature version of the BRST structure identified in \cite{Crossley:2015evo} and can be recovered from \eqref{eq:skkmsalg} provided we set $\Fs_{\thb\thb} = \Fs_{\theta\theta} =0$ and $\Fs_{\theta\thb} = -i$.

A hydrodynamic effective action ought to capture the response functions, and the attendant hydrodynamic fluctuations. To this end, note that hydrodynamic variables capture the universal low energy dynamics in near-equilibrium states for any interacting quantum system. The pion fields of hydrodynamics  are maps $X^\mu(\sigma^a)$ from a reference worldvolume to the physical spacetime, such that the push-forward of a reference thermal vector gives the physical thermal vector (this is a modern version of the ``Lagrangian'' description of a fluid). They may be viewed as breaking the individual diffeomorphisms of the SK construction to the diagonal \cite{Nickel:2010pr}. Incorporation of the BRST symmetries can be done easily by passing to a superspace with fields: 
\begin{equation}
\SF{X}^\mu = X^\mu + \theta \, \xpsib^\mu + \thb\, \xpsi^\mu + \thb\theta \left(\tx^\mu -\Gamma^\mu_{\rho \sigma} \,\xpsib^\rho \, \xpsi^\sigma\right).
\label{eq:xsf0}
\end{equation}	
The bottom component of this superfield,  $X^\mu$, is the classical hydrodynamic field and the physical thermal vector is the push-forward of the background thermal vector, $\Kref^\mu = \SF{\Kref}^I \Dut_I \SF{X}^\mu |$. The top component  is the fluctuation field $\tx^\mu$, while $\xpsi^\mu$ and $\xpsib^\mu$ are the BRST ghosts. The superfield $\SF{X}^\mu$ carries $\UT$ charge and transforms as 
\begin{equation}
\SF{X}^\mu  \mapsto \SF{X}^\mu + \SF{\Lambda}\, \lieD_\Kref \SF{X}^\mu \,.
\label{eq:xut}
\end{equation}	
\begin{figure}[ht!]
\begin{center}
\includegraphics[width=0.48\textwidth]{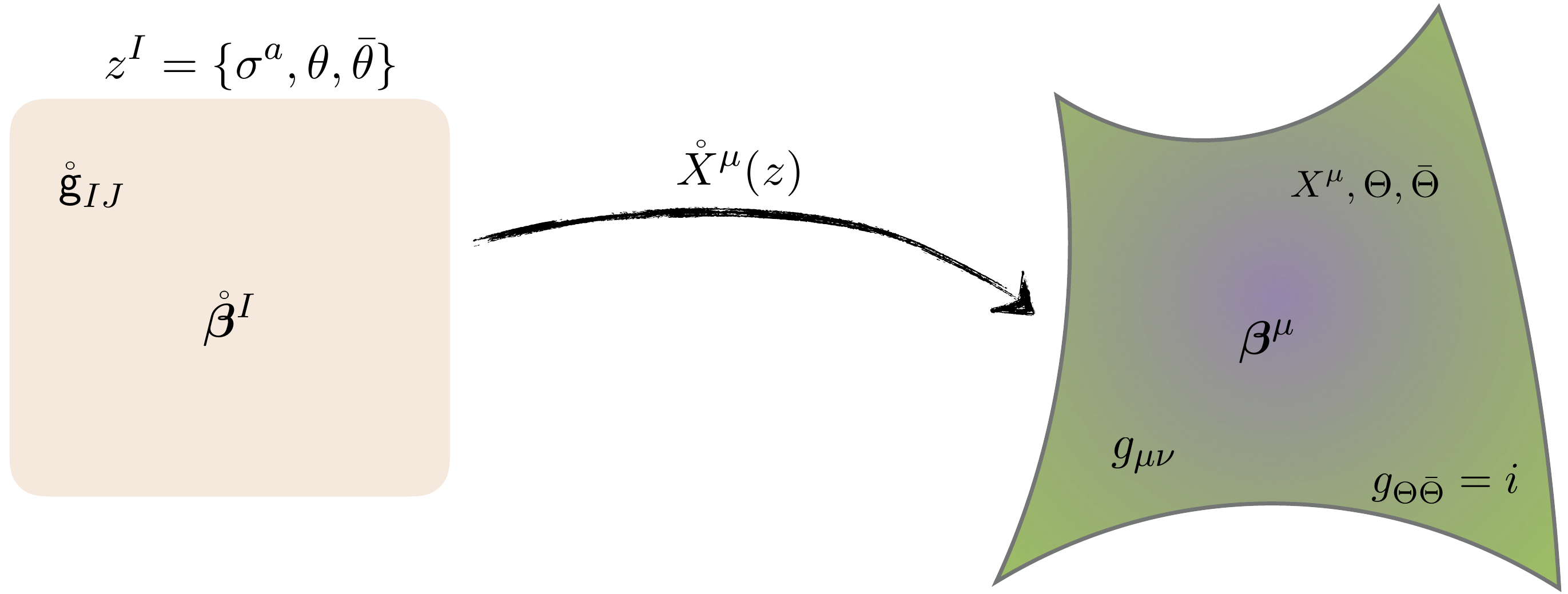}
\caption{Illustration of the data for hydrodynamic sigma models. The physical degrees of freedom are captured in the target space maps $\SF{X}^\mu(\sigma^a,\theta,\thb)$. The worldvolume geometry is equipped with a rigid super-vector field $\SF{\Kref}^I$, which pushes forward to the physical thermal vector in spacetime, while the spacetime metric $g_{\mu\nu}, \;g_{\Theta \bar{\Theta}} = -g_{\bar{\Theta}\Theta } =i $, pulls back to the worldvolume metric $\SF{\gref}_{IJ}$. The target space Grassmann coordinates are gauge fixed to be aligned with the worldvolume ones, $\Theta = \theta$, 
$\bar{\Theta} =\thb$. }
\label{fig:sigmamap2}
\end{center}
\end{figure}

The effective field theory is further constrained by the discrete,  anti-linear ${\mathbb Z}_2$ transformation: {\sf CPT} exchanges $\thetab\leftrightarrow\theta$ and hence acts as an R-parity on the superspace and acts as usual on the coordinates $\sigma^a$. The action can be intuited from using the fact that the $\thetab\theta$ component of superfields contains the difference operators.  Of crucial import is the observation that the gauge sector \eqref{eq:Aform} contains only one covariant field of ghost number zero, $\Fs_{\theta\thb}|$. We assume that the topological gauge dynamics is consistent with the existence  of a vacuum where $\vev{\Fs_{\theta\thb}|} = -i$, which spontaneously breaks {\sf CPT}. This symmetry breaking pattern would then explain the emergence of entropy and the associated arrow of time, with $\vev{\Fs_{\theta\thb}|}$ being the order parameter for dissipation.

In fact, much of the discussion below could naively be understood by simply working with a background $\UT$ gauge field $\As_I$, restricted to ensure that the field strength component $\vev{\Fs_{\theta\thb}|} = -i$, {\sf CPT} is broken explicitly and the inflow is biased towards entropy production. We refer to this limit as the MMO limit after Mallick-Moshe-Orland \cite{Mallick:2010su}.  On physical grounds, since we expect {\sf CPT} breaking to emerge dynamically rather than being imposed from the beginning, we expect a complete theory to involve the dynamics of $\Fs_{\theta\thb}$ as well as 
saddle points where $\vev{\Fs_{\theta\thb}|} \neq -i$. Indeed given the  conservation of entropy in superspace, there does not seem to be  any fundamental obstructions to gauging the thermal diffeomorphisms and making the superspace gauge fields dynamical.

\section{Dissipative hydrodynamic actions and entropy inflow}
\label{sec:}

Once we have identified the symmetries and the basic fields, we can proceed to construct an effective action, which has to be a topological sigma model. The basic variables are $\SF{X}^\mu$ (Grassmann coordinates are fixed to be $\Theta  = \theta, \bar{\Theta} = \thb$)  but they cannot appear in their bare form, owing to target space diffeomorphism invariance. We pull back the target space metric (with $g_{\Theta \bar{\Theta}} = i, g_{\mu\Theta} = g_{\mu{\bar\Theta}} =0$) to a worldvolume metric 
\begin{align}
\SF{\gref}_{IJ}(z) = g_{\mu\nu}(\SF{X}(z))\,\Dut_I \SF{X}^\mu\,\Dut_J \SF{X}^\nu  + g_{\Theta \bar{\Theta}} \left( 
\Dut_I \SF{\Theta} \, \Dut_J \, \SF{\bar{\Theta}}  - \Dut_I \SF{\bar{\Theta}} \, \Dut_J \, \SF{\Theta}\right) ,
\label{eq:gwv}
\end{align}
 and use it to build an effective action. This pullback induces the worldvolume measure $\frac{\sqrt{-\SF{\gref}}}{\zsf}$ with $\zsf$ defined below in Eq.~\eqref{eq:fullLag}  (see \cite{Haehl:2015uoc}).
 We also introduce a worldvolume covariant super-derivative $\Dwv_I$ which differs from the $\UT$ covariant derivative $\Dut_I$ by terms involving $\Fs_{IJ}$ and $\SF{\Kref}^I$ when acting on worldvolume tensors, such as to ensure covariance with respect to worldvolume diffeomorphisms. The precise definition of $\Dwv_I$ will not be material below; all we require is the fact that $\Dwv_I$ allows for integration by parts in superspace.\footnote{ In practice this only requires that $\Dwv_I$ be compatible with the $\UT$ covariant pullback measure   $\Dwv_I\left(\frac{\sqrt{-\SF{\gref}}}{\zsf}\right) =0$, which is weaker than being metric compatible as is usually assumed.  We demonstrate the existence of such a derivation in \cite{Haehl:2018ab}. } 
To write an effective action, we begin by writing down the theory capturing the trivial topological sector in terms of the fields introduced so far. We then get physical hydrodynamics by breaking this topological limit by introducing a background source  $h_{IJ}$ for energy-momentum tensor
($\SF{\gref}_{IJ} \mapsto \SF{\gref}_{IJ} + \thb\theta\, h_{IJ}$) and $\mathcal{F}_I$ for free energy current ($\As_I \mapsto \As_I + \thb\theta \, \mathcal{F}_I$).

Hydrodynamic transport consistent with the second law is described by effective actions of the form 
\cite{Haehl:2015uoc,Haehl:2018ab}:
\begin{equation}
S_\text{wv} = \int d^d \sigma \,  d\theta\, d\thetab\,\frac{\sqrt{-\SF{\gref}}}{\zsf} \; \SF{\Lagref} \,,\qquad \zsf \equiv 1 + \As_I \SF{\Kref}^I\,,
\label{eq:fullLag}
\end{equation}
where $\SF{\Lagref}$ is a scalar functional of the worldvolume fields, invariant under the following symmetries: $(i)$ ordinary spacetime super-diffeomorphisms, $(ii)$ reparameterizations $z^I \mapsto z^I + f^I(\sigma^a)$ of the worldvolume super-coordinates, $(iii)$ an anti-linear {\sf CPT} involution, $(iv)$ ghost number conservation. Invariance under $(v)$ the BRST symmetry and $\UT$ is manifest in the superspace formulation. We define the energy-momentum and free-energy  Noether super-currents:
\begin{equation}
\begin{split}
\SF{\TEMref}_{}^{IJ} 
&\equiv 
	\frac{2}{\sqrt{-\SF{\gref}}} \, \frac{\delta}{\delta \SF{\gref}_{IJ}}
	\prn{\sqrt{-\SF{\gref}}\ \SF{\Lagref}}\,,
\\ 
\SF{\Nref}_{}^I 
&\equiv
	 -\frac{\zsf}{\sqrt{-\SF{\gref}}} \, \frac{\delta }{\delta \SF{\Ascr}_{I}}
	 \prn{\frac{\sqrt{-\SF{\gref}}}{\zsf}\ \SF{\Lagref} } \,. 
\end{split}
\label{eq:TNwvdef}
\end{equation}	
Below we will present abstract lessons to be learned from the superspace effective action, especially regarding dissipation. In \cite{Haehl:2018ab} we will show explicitly that the  general action of the form \eqref{eq:fullLag} consistent with the symmetries mostly reproduces the eightfold classification of ordinary space currents $\{\TEMref^{ab},\Nref^a\}$.\footnote{In the present analysis we postpone a discussion of both charged and anomaly induced transport, which require some additional notation and formalism. Corresponding generalizations along the lines of \cite{Haehl:2013hoa,Haehl:2015pja} are expected to be straightforward.} Note that these will generically contain fluctuation fields (terms involving $\tilde{X}^\mu$). The theory thus not only gives classical hydrodynamic constitutive relations, but also predicts the form of statistical fluctuations. Appendix \ref{sec:bulkvisc} provides a simple example illustrating the formal manipulations described in this section.

Physical equations of motion follow from varying the action with respect to the dynamical pullback superfield $\SF{X}^\mu$. This enters the action through the metric \eqref{eq:gwv} and therefore leads to energy-momentum conservation in superspace:
\begin{equation} 
\delta_{\SF{X}} S_\text{wv} = \int d^d \sigma d\theta d\thb\, \frac{\sqrt{-\SF{\gref}}}{\zsf} \;  
\Dwv_I \left( \SF{\TEMref}^{IJ} \, \Dwv_J \SF{X}^\mu  \right)g_{\mu\nu}\, \delta \SF{X}^\nu .
\label{eq:supereom}
\end{equation}
While this equation of motion involves all the super-components of the energy-momentum tensor, one can demonstrate equivalence to the classical hydrodynamic equations by noting that  it produces correct target space dynamics after projecting down to the bottom component up to fluctuations and ghost bilinears and pushing forward using $X^\mu$: 
\begin{equation}
\begin{split}
0 &=\Dwv_I \left( \SF{\TEMref}^{IJ} \, \Dwv_J \SF{X}^\nu \right)  \big| \\
&  = \nabla_\mu T^{\mu\nu} + \text{ghost bilinears + fluctuations.}
\end{split}
\label{eq:}
\end{equation}	

For the analogue of the adiabaticity equation \eqref{eq:ordadiab}, consider a $\UT$ transformation by a gauge parameter $\SF{\Lambda}$ which transforms the action via
\begin{equation}\label{eq:utcalcfull}
\begin{split}
\delta_{\SF{\Lambda}}   S_\text{wv} 
&=
	 \int d^d \sigma d\theta d\thetab\, \frac{\sqrt{-\SF{\gref}}}{\zsf}  
	 \left\{\frac{1}{2}  \SF{\TEMref}_{}^{IJ}  \, (\SF{\Lambda}, \SF{\gref}_{IJ})_\Kref  +( \Dwv_I \SF{\Lambda})\,  \SF{\Nref}_{}^I   \right\}\\
&=
	\int d^d \sigma d\theta d\thetab\, \frac{\sqrt{-\SF{\gref}}}{\zsf} \; \SF{\Lambda} \,\left\{ \frac{1}{2} \, \SF{\TEMref}_{}^{IJ}  \, \lieD_\Kref \SF{\gref}_{IJ}  -  \Dwv_I  \SF{\Nref}_{}^I   \right\} \,.
\end{split}
\end{equation}
We have  invoked the $\UT$ transformation of the pullback metric \eqref{eq:gwv}, inherited from \eqref{eq:xut}, and integrated  by parts in superspace. Invariance then implies the \emph{super-adiabaticity equation}:
\begin{equation}\label{eq:superadiabatic} 
\Dwv_I  \SF{\Nref}_{}^I-\frac{1}{2} \, \SF{\TEMref}_{}^{IJ}  \, \lieD_\Kref \SF{\gref}_{IJ}    =0  \,.
\end{equation}
This equation is a Bianchi identity for the $\UT$ transformation and consequently is insensitive to the gauge dynamics.

The $\UT$ super-adiabaticity equation \eqref{eq:superadiabatic} can be decomposed into ordinary and superspace contributions:\footnote{ Some signs may seem counterintuitive but are a consequence of Grassmann odd index contractions; we use DeWitt conventions  \cite{DeWitt:1992cy}.}
\begin{widetext}
\begin{equation}
\begin{split}
\underbrace{\left(\Dwv_a \SF{\Nref}^a 
-\frac{1}{2} \, \SF{\TEMref}^{ab}\,\lieD_\Kref\, \SF{\gref}_{ab}\right)\Big| }_{\text{classical + fluctuations}} 
= -  \underbrace{ \left( \Dwv_\theta \SF{\Nref}^\theta + \Dwv_\thb \SF{\Nref}^\thb + \SF{\TEMref}^{\theta \thb }\,\lieD_\Kref\, \SF{\gref}_{\theta \thb} +\SF{\TEMref}^{a\theta}\,\lieD_\Kref\, \SF{\gref}_{a\theta}
+ \SF{\TEMref}^{a\thb}\,\lieD_\Kref\, \SF{\gref}_{a\thb}
 \right)\Big|}_{\text{entropy inflow}} \,.
\end{split}
\label{eq:superadcomp}
\end{equation}
\end{widetext}
Modulo contributions from the fluctuation fields (and ghost bilinears), the terms we have isolated on the l.h.s., are precisely the combination appearing on the l.h.s.\ of \eqref{eq:ordadiab} (now in the worldvolume theory instead of physical spacetime). This naturally suggests interpreting the second set of terms indicated as `entropy inflow' as the part that contributes to the total entropy production $\Delta$. Hence,  we identify the r.h.s.\ of \eqref{eq:superadcomp} as
\begin{equation}
\begin{split}
\Delta 
& =-\left( \mathfrak{D}_\theta \Nref^\theta+  \mathfrak{D}_\thb \Nref^\thb 
  	 \right)  \big|+ \text{ghost bilinears}\,,
\label{eq:Delta0}
\end{split}
\end{equation}	
%
where we have used the conserved ghost number charge to identify the additional terms as being built out of ghost bilinears (note also that $\lieD_\Kref\, \SF{\gref}_{\theta \thb} |=0$). We interpret this equation as saying that the entropy production is controlled by the Grassmann-odd descendants of the free energy and energy-momentum super-tensors.\footnote{ The bosonic part  $\Delta_{bos} = -\big( \mathfrak{D}_\theta \Nref^\theta+  \mathfrak{D}_\thb \Nref^\thb \big) \big|$ can be obtained directly from the effective action by turning on a suitable source $\sBdel$, cf., Appendix \ref{sec:bulkvisc} for an example.
\label{foot:sBdel}}

Let us take stock: $\UT$ invariance implies a super-Bianchi identity which, suitably split up, provides a rationale for entropy production. In particular, the superspace components of the free energy current are primarily responsible for entropy production. This is highly reminiscent of the inflow paradigm for 't Hooft anomalies \cite{Callan:1984sa}. The basic idea there is that the effective action for an anomalous symmetry can be presented in a covariant form by viewing the anomalous field theory as the edge/boundary dynamics of a bulk topological field theory. This is familiar from the classic example of quantum Hall edge states (equivalently, the Chern-Simons, WZW correspondence). Our proposition is that entropy production can be viewed in an identical manner in the $\UT$ covariant formalism, with the inflow occurring from Grassmann odd directions into the physical degrees of freedom. Despite $\{\Nref^\theta,\Nref^\thb\}$ being ghost currents, their derivatives occurring in \eqref{eq:Delta0} are bosonic and physical, and capture entropy production.

Our final task is to explain why $\Delta \geq 0$, i.e., the second law. One can in principle argue abstractly by invoking the Jarzynski relation \cite{Jarzynski:1997aa,Jarzynski:1997ab} which follows from the superspace effective action once we assume a {\sf CPT} breaking pattern \cite{Mallick:2010su,Gaspard:2012la,Haehl:2015uoc}. It is more instructive to see this directly by relating to Bhattacharyya's theorem. To this end we recall that the effective action for dissipative transport is given by \cite{Haehl:2015uoc}
\begin{equation}
S_\text{wv,diss} = \int d^d\sigma d\theta d\thb\; \frac{\sqrt{-\SF{\gref}}}{\zsf}\,\left(-\frac{i}{4}\right) \, \SF{\etaref}^{IJKL} \, \gpsib_{IJ} \, \gpsi_{KL} \,, \qquad 
 \gpsib_{IJ} \equiv \Dut_\theta \SF{\gref}_{IJ}\,, \;\;
\gpsi_{KL} \equiv  \Dut_\thb\SF{\gref}_{KL}  \,.
\label{eq:Ldiss}
\end{equation}	
At first sight, the tensors  ${\gpsi}_{IJ} ,{\gpsib}_{KL}$ may appear non-covariant and hence forbidden. However, as is easily verified, the worldvolume diffeomorphisms specified as symmetry $(ii)$ above are such that these terms are in fact covariant and hence allowed (further details may be found in \cite{Haehl:2018ab}).\footnote{Note that under reparameterizations $z^I \mapsto z^I + f^I(\sigma^a)$ the worldvolume connection components $\Cwv^{I}_{\ \theta J}$ and $\Cwv^{I}_{\ \thb J}$ (and hence $ \gpsib_{IJ} \equiv \Dwv_\theta \SF{\gref}_{IJ} + (-)^{IK} \, \Cwv^K_{\ \theta I} \SF{\gref}_{KJ} +(-)^{K(I+J)+IJ} \, \Cwv^K_{\ \theta J} \SF{\gref}_{IK} $ etc.) are covariant.} 

The dissipative tensor $\SF{\etaref}^{IJKL}$  satisfies: 
\begin{equation}
\begin{split}
& \SF{\etaref}^{(IJ)(KL)} = (-)^{IJ}\; \SF{\etaref}^{(JI)(KL)} = (-)^{KL}\,\; \SF{\etaref}^{(IJ)(LK)}  \\
& \SF{\etaref}^{(IJ)(KL)}  = (-)^{(I+J)(K+L)}\; [ \SF{\etaref}^{(KL)(IJ)} ]^{\text{\sf{\tiny CPT}}}\end{split}
\label{eq:etacpt1}
\end{equation} 
but is an otherwise arbitrary covariant functional of the hydrodynamic worldvolume fields (see Appendix \ref{sec:bulkvisc} for an example).
Direct computation of the currents leads to:
\begin{equation}
\begin{split}
\Delta 
&=
	\frac{i}{4}\, (\Fs_{\theta\thb}| ) \, \etaref^{abcd}\, \lieD_\Kref \gref_{ab}\; \lieD_\Kref \gref_{cd}  + \text{fluctuations} + \text{ghost bilinears} \,.
\end{split}
\label{eq:Delgg}
\end{equation}	
We now invoke the expectation value $\vev{\Fs_{\theta\thb}|} = -i $, thus spontaneously breaking {\sf CPT} as the origin of dissipation \cite{Haehl:2015uoc}. As long as $\etaref^{abcd}$ is a positive definite map from the space of symmetric two-tensors to symmetric two-tensors, the amount of entropy produced is then non-negative definite, viz., $\Delta \geq 0$, which is precisely the condition established earlier.
The easiest way to argue for the positivity of $\etaref^{abcd}$ is by performing the superspace integral in the dissipative effective action which results in an imaginary part of the form $\text{Im}(S_\text{wv,diss}) = \frac{1}{4}\int d^d \sigma \sqrt{-\gref} \; \etaref^{abcd} h_{ab} h_{cd}$. Thus, in order for the work done by the sources to be compatible with a convergent path integral, we need to enforce positivity on $\etaref$ at leading order in the derivative expansion (see \cite{Glorioso:2016gsa,Haehl:2015uoc,Haehl:2018ab}).

\medskip
 \noindent
 \emph{\it{Note added:}} The work described here was first presented at the ``It from Qubit workshop'', Bariloche  (Jan 2018) \cite{Bariloche:2018aa}. While this manuscript was under preparation we  received \cite{Jensen:2018hhx} which overlaps with the material discussed here.

\begin{acknowledgments}

It is a pleasure to thank  Shiraz Minwalla,  Veronika Hubeny and Spenta Wadia for discussions on the origins of entropy production in hydrodynamics and beyond.
FH gratefully acknowledges support through a fellowship by the Simons Collaboration `It from Qubit'. RL gratefully acknowledges support from International Centre for Theoretical Sciences (ICTS), Tata institute of fundamental research, Bengaluru. RL would also like to acknowledge his debt to all those who have generously supported and encouraged the pursuit of science in India. MR is supported in part by U.S.\ Department of Energy grant DE-SC0009999 and by the University of California. MR would like to thank TIFR, Mumbai and ICTS, Bengaluru, as well as the organizers of the ``It from Qubit workshop'' at Bariloche, and  ``Chaos and Dynamics in Correlated Quantum Matter'' at Max Planck Institute for the Physics of Complex Systems, Dresden for hospitality during the course of this work.

\end{acknowledgments}


\newpage
\appendix

\begin{widetext}
\section{Example: bulk viscosity} 
\label{sec:bulkvisc}

In this section we give an explicit example of a transport coefficient in order to illustrate the abstract ideas introduced in this letter. The example we will consider is bulk viscosity $\zeta(T)$. 

The superspace Lagrangian for bulk viscosity is 
\begin{equation}
\SF{\Lagref}^{^{(\zeta)}} = 
	 - \frac{i}{4}\, \zeta(\SF{T}) \,\SF{T} \, \SF{P}^{IJ}\SF{P}^{KL} \,  \gpsib_{IJ}  \,
	  \gpsi_{KL} \,,
\label{eq:Lageta}
\end{equation}
where $\SF{P}^{IJ} = \SF{T}^2 \SF{\Kref}^I \SF{\Kref}^J + \SF{\gref}^{IJ}$. This is, of course, an example of the structure \eqref{eq:Ldiss}.
We can explicitly perform the superspace expansion of \eqref{eq:Lageta}. For simplicity, we only keep track of ghost number zero fields and sources. This yields:
{
\begin{equation}
\begin{split}	
S_{\text{wv}}^{^{(\zeta)}} &=\int d^d\sigma\, d\theta\, d \thb \;  \SF{\Lagref}^{^{(\zeta)}}   \\
&=
	 \int d^d\sigma\, \sqrt{-\gref} \,\left(- \frac{i}{4}\right) \, T\, \zeta(T)  
	 \bigg[ -i \, P^{ab} \prn{ \source{h_{ab}} + \tilde{\gref}_{ab} -
 	  \big((\Fs_{\theta\thb}|) -\sBdel, \gref_{ab}\big)_\Kref }+2 \source{h_{\theta\thb}}  \bigg] \\
& \qquad\qquad\qquad\qquad  \times \bigg[ -i \, P^{ab} \prn{ \source{h_{ab}} + \tilde{\gref}_{ab} +
 	  ( \sBdel, \gref_{ab})_\Kref }  +2 \source{h_{\theta\thb}}  \bigg] \,.
\end{split}
\label{eq:bvterm3}
\end{equation}
}\normalsize
where $\tilde{\gref}_{ab}= \partial_\theta \partial_\thb \SF{\gref}_{ab} |$. In the above we turned on a source 
$\sBdel$ which enters via shifts in the $\UT$ gauge field $\As_I$: $\As_\theta \mapsto \As_\theta -\thb\, \sBdel$ and $\As_\thb \mapsto \As_\thb + \theta\, \sBdel$. Variation with respect to $\sBdel$ directly yields the entropy production $\Delta$.
This ordinary space action captures the contributions to the bulk viscosity in the effective action, including the entropy current and fluctuation terms. We can read off the constitutive relations according to \eqref{eq:TNwvdef}:
\begin{equation}
\begin{split}
{\TEMref}_{_{(\zeta)}}^{ab} 
&\equiv 
	\frac{2}{\sqrt{-{\gref}}} \, \frac{\delta S^{^{(\zeta)}}_\text{wv}}{\delta \source{h_{ab}}} 
	= \underbrace{ -i(\Fs_{\theta\thb}|) \, \zeta(T) \, \vartheta \, P^{ab}  }_{\text{classical}} \,
	  	+ \underbrace{\,i\,T \, \zeta(T) \,P^{ab} \, P^{cd} \, \tilde{\gref}_{cd} }_{\text{fluctuations}}
\,,\qquad\quad
{\Nref}_{_{(\zeta)}}^a
\equiv
	- \frac{1}{\sqrt{-{\gref}}} \, \frac{\delta S^{^{(\zeta)}}_\text{wv}}{\delta \sAt{a}}
 =  0  
 \,.
\end{split}
\label{eq:TNwvdefideal}
\end{equation}	
where we defined the expansion $\vartheta = \nabla_a u^a$. 
The classical current takes precisely the form of bulk viscosity transport (after setting $\langle \Fs_{\theta\thb}| \rangle = -i$). We can combine the classical and the fluctuation parts in the suggestive form
\begin{equation}
   {\TEMref}_{_{(\zeta)}}^{ab}  = - \frac{1}{2} \, T \, \zeta(T) \, P^{ab} \, P^{cd} \, \prn{\lieD_\Kref\gref_{cd}  - 2i \, \tilde{\gref}_{cd} } \,.
\end{equation}
The equation of motion follows by varying the action with respect to $\tilde{X}^\mu$. To do this explicitly, note that $\tilde{X}^\mu$ appears inside $\tilde{\gref}_{ab}$ in the form $\tilde{\gref}_{ab} \supset 2 \, g_{\mu\nu} \, \partial_{(a} X^\mu \partial_{b)} \tilde{X}^\nu + \tilde{X}^\rho \, \partial_\rho g_{\mu\nu}  \, \partial_a X^\mu \,\partial_b X^\nu$. It is easy to verify that variation with respect to $\tilde{X}^\mu$ therefore yields 
\begin{equation}
\nabla_\mu T^{\mu\nu}_{_{(\zeta)}} = 0\,,
\end{equation} 
 where $T^{\mu\nu}_{_{(\zeta)}} = {\TEMref}_{_{(\zeta)}}^{ab} \, \partial_a X^\mu \partial_b X^\nu$ is the push-forward to physical spacetime.

The full entropy production can also be obtained from the effective action by a variation with respect to the source $\sBdel$:
\begin{equation}
\begin{split}
 \Delta_{_{(\zeta)}} &\equiv
	 -\frac{1}{\sqrt{-{\gref}}} \frac{\delta S^{^{(\zeta)}}_\text{wv}}{\delta \sBdel}   
	  	 = \partial_a \Nref_{_{(\zeta)}}^a - \frac{1}{2} \, {\TEMref}_{_{(\zeta)}}^{ab} \, \lieD_\Kref \gref_{ab} 
	=  \underbrace{i(\Fs_{\theta\thb}|) \,\frac{\zeta}{T} \, \vartheta^2}_{\text{classical}} - \underbrace{\frac{i}{2} \, T \, \zeta\, P^{ab} P^{cd} \, \lieD_\Kref \gref_{ab} \, \tilde{\gref}_{cd}}_{\text{fluctuations}}  \,,
 \end{split}
\end{equation}
which can be verified explicitly. Again, after setting $\langle \Fs_{\theta\thb}| \rangle = -i$, the classical part of entropy production matches the expectation for bulk viscosity \cite{Haehl:2015pja}. The fluctuation part is a prediction of our formalism. 
The second law constraint is the requirement that $ \Delta^\text{(classical)}_{_{(\zeta)}} \geq 0$, which clearly translates to the well-known $\zeta \geq 0$. 
Similar analysis can be performed for the shear viscosity and  corresponding fluctuation terms derived (see \cite{Haehl:2018ab}).

\end{widetext}

\end{document}